\documentclass[conference]{IEEEtran}
\usepackage{cite}

\ifCLASSINFOpdf
  \usepackage[pdftex]{graphicx}
\else
\fi

\begin{document}
%
\title{An Initial Study on Ideal GUI Test Case Replayability}

\author{\IEEEauthorblockN{Arthur-Jozsef Molnar}
\IEEEauthorblockA{Department of Computer Science\\ Faculty of Mathematics and Computer Science\\ Babe\c{s}-Bolyai University}
}

\maketitle

\begin{abstract}
In this paper we investigate the effect of long-term GUI changes occurring during application development on the reusability of existing GUI test cases. We conduct an empirical evaluation on two complex, open-source GUI-driven applications for which we generate test cases of various lengths. We then assess the replayability of generated test cases using simulation on newer versions of the target applications and partition them according to the type of repairing change required for their reuse.
\end{abstract}

\IEEEpeerreviewmaketitle

\section{Introduction}

Many software applications today employ graphical user interfaces (GUIs) to interact with users. As a highly successful paradigm, we encounter GUI-driven applications on many types of devices, a trend that seems set to continue in today's world of pervasive computing. However, while these applications are ubiquitous, the same thing cannot be said about the processes that should support their life cycle, such as quality assurance (QA). Because as much as 50\% of application code can be GUI related \cite{20}, the existence of QA processes for these applications becomes crucial. Recent work on GUI testing consists of notable contributions such as developing theoretical frameworks for GUI testing \cite{20,93,94}, implementing advanced testing tools \cite{95,102}, studying possibilities for automation \cite{96,142} and gathering empirical evidence regarding the success of automated testing of complex applications \cite{66,67,96}.

Tools for GUI testing can be divided into two categories: capture-replay and model-based. 

Capture-replay tools \cite{146} such as Pounder\footnote{http://pounder.sourceforge.net}, Marathon\footnote{http://www.marathontesting.com} or jfcUnit\footnote{http://jfcunit.sourceforge.net} are generally representative of the first wave of automated tooling and work in the two phases that spawned their name: during the first capture phase, the tester works with the application under test (AUT) and manually records test cases which are stored by the tool and then replayed during the second phase. This approach introduces the GUI paradigm to the generation of test cases by allowing them to be built by interacting with the AUT. However, typical capture-replay tools suffer limitations when the behaviour or GUI of the tested systems change. Also, it must be noted that such tools are only able to automate one part of the testing process, as creating test cases remains an overwhelmingly manual undertaking. Also, typical capture-replay tools cannot provide comprehensive oracles for GUI testing \cite{15} beyond crash-detection and recognizing error windows. These issues were well known to practitioners and tool developers so many ideas were implemented to alleviate such limitations. 

In \cite{17}, Takahashi proposes intercepting Win32 API graphic calls to replace screen captures, as they are more reliable and occupy less storage space when persisted. Another approach is described in \cite{90}, where Ostrand et al. design a Test Development Environment (TDE) that links a test designer and a test generation library with a standard capture-replay tool. Its feasibility is then tested by generating test cases for a medical diagnosis machine.

Many of the shortcomings of capture-replay tools can be addressed using model based approaches. The availability of a model allows automated generation and execution of test cases and helps with implementing necessary test oracles to evaluate testing results. The typical drawback is represented by the time and effort dispensed for model building and validation, as a suitable balance must be achieved between model complexity and system testability to enable revealing system faults \cite{194}. Recent advances addressing such issues come from Microsoft Research's NModel, that uses C\# for building the model \cite{195} and Silva et al. \cite{153} who employ Spec Explorer for GUI testing.

A solid body of research in model-based GUI testing\footnote{Literature refers to \emph{GUI testing} as testing an application through its GUI.} was initiated by Memon's PhD thesis \cite{20}. He provides a definition for the state of the GUI (\cite{20}, p.29), the event-flow graph (\cite{20}, p.37) which models the valid flow of GUI events and for the GUI test case (\cite{20}, p.42). The theoretical foundations are then used for implementing the GUITAR\footnote{http://guitar.cs.umd.edu} GUI testing framework used in the presented research.  

One of the outstanding issues in automated testing regards test case maintenance. Like all application layers, GUIs invariably change during application development and maintenance. Widgets can be resized, moved or changed in numerous ways to fit new application requirements. This leads to many test cases becoming unusable on newer versions of the target applications as testing tools cannot recognize the changed GUI elements. This constitutes a major hurdle in the automation of the process, regardless of approach.

Our research aims to assess how typical changes in GUI-driven applications affect the reusability of existing test cases. In this sense, we study an idealized situation where we categorize existing test cases according to their degree of replayability using known correct information obtained by studying GUI changes occurring in two complex, open-source GUI applications.

The structure of this paper is as follows: the second section introduces required preliminaries and related work. The third section details the target applications used in our research and presents our initial case study which provides both cross-sectional and longitudinal perspectives. The fourth section overviews threats to the validity of our empirical evaluation while the last section is reserved for conclusions and future work planned.

\section{Preliminaries}

In this section we describe related work regarding GUI test case maintenance and we briefly describe the GUITAR GUI testing framework that we extensively use in our case study.

\subsection{Related work}

The problem of GUI test case maintenance has not gone unnoticed and several approaches have been proposed. In \cite{97}, Memon proposes \emph{repairing transformations} that insert and remove test case steps with the aim of repairing tests broken by changes in the AUT. An empirical evaluation is then performed where the efficiency of the proposed process is evaluated using four open source applications, one of which is the FreeMind mind-mapper also employed in our research. Huang et al. use genetic algorithms to repair automatically generated infeasible test cases in \cite{98}. In their study they use the same theoretical foundation as \cite{97} and evaluate obtained results on several synthetic applications. These approaches prove that GUI test cases can be successfully repaired to run on modified versions of the AUT. However, they are of limited use in long-term regression testing because by altering the sequence of test steps they do not replay \emph{exactly} the same test steps on the modified application. McMaster and Memon detail preliminary work in enabling regression testing of GUI applications in \cite{113}, where they describe a conceptual heuristic process to find functionally equivalent widgets across versions of a GUI application. In our previous research we implemented such a process and showed it achieves high accuracy in correctly classifying GUI elements across many application versions \cite{177}.

However, because all previous approaches to repair test cases are subject to error, we were interested in performing an evaluation on the efficiency of an ideal error-free process when employed for long-term GUI test case maintenance in the case of complex GUI applications. Such an evaluation provides a benchmark against which existing and future implementations can be compared and is useful for assessing how many test cases can be replayed when using perfect\footnote{Repaired test cases are as functionally close to the original as possible.} implementations. Such a perfect implementation can be obtained by formally documenting GUI changes, annotating GUI elements so they are recognized across versions by the testing harness or approximated using highly accurate heuristic processes.

\subsection{The GUITAR framework}

The theoretical aspects presented in \cite{20} were implemented in the GUITAR testing framework \cite{95,96}. GUITAR is a mature testing toolset that automates many processes in testing: obtaining the GUI model, generating valid test cases, replaying them on the AUT and recording information usable by test oracles. Therefore it facilitates the adoption of model based testing by decreasing the effort of obtaining the GUI model and providing automation for associated activities. GUITAR's components are available\footnote{As of February, 2012} for Java and Web, with Windows, Android and iOS implementations currently in development. This makes GUITAR a state of the art tool in the research and practice of GUI application testing. Its four main components, briefly described in the order in which they are usually employed are:
\begin{itemize}
\item\emph{GUIRipper} can be used to automatically record the AUT's GUI model in XML format. Using reflection and automated interaction, it records all accessible application windows together with their widgets and properties \cite{95}.
\item\emph{GUI2EFG} takes as input the model obtained using the GUIRipper and computes the application's event-flow-graph that provides the valid event sequences within the GUI of the AUT \cite{141}. This component provides crucial functionality for building valid test cases, as not every widget is actionable at all times.
\item\emph{TestCaseGenerator} can build valid test cases using the provided GUI model and event-flow-graph. The \emph{TestCaseGenerator} uses a plugin architecture that allows implementing new strategies for generating test cases \cite{142}.
\item\emph{TestCaseReplayer} is used to run generated test cases and record the target GUI state after each test step, allowing offline analysis of test case execution. The \emph{Replayer} component was used in several studies. In \cite{96} authors use it for research in regression testing, while in \cite{101} Brooks and Memon employ it for automating profile-guided testing. In \cite{144}, Xie and Memon use GUITAR to study desirable characteristics of GUI test suites while in \cite{145} a pilot study assesses GUI event interactions and the influence of event context on test case outcome.
\end{itemize}
Some of GUITAR's limitations stem from the theoretical foundations from which it was developed \cite{20} and regard testing GUIs that present continuous streams of data or that interface real-time systems. Due to their particular constraints, they might not be handled properly by GUITAR's components and are therefore not targeted by our research. More so, to the best of our knowledge there exist no readily-available tools for assisting QA processes targetted towards such GUIs, leaving GUITAR as the prime candidate for carrying out our empirical investigation. In our research we use \emph{GUIRipper} to obtain the GUI models of our target applications, which we then pass through \emph{GUI2EFG} to compute the valid event sequences. Finally, we use GUITAR's test generation component to generate the test cases used in our evaluation.

\section{Case Study}

In this section we present our initial case study in which we investigate how changes that occur during application development affect replayability of existing GUI test cases. As GUI test case steps action widgets, assessing test case replayability requires information about the functionally equivalent widgets \cite{113} between the studied versions. This information was obtained using an automated heuristic process \cite{177}, with all results manually double-checked for correctness. The amount of effort involved limited our study to two target applications: the FreeMind\footnote{http://freemind.sourceforge.net/wiki/index.php/Main\_Page} mind-mapper and the jEdit\footnote{http://jedit.org} text editor, both detailed in the following section. Using their publicly available source code repositories we downloaded 30 distinct versions of these applications to obtain a suitable balance between generality and the amount of effort required to prepare the data. Next, we generated comprehensive test suites for each version using GUITAR's test generator. One of our goals was assessing the effect of GUI changes on test cases of different lengths. Therefore, for each application version we generated a test suite in the following manner:
\begin{itemize}
\item \emph{Event-interaction coverage\footnote{All valid length-2 test cases.}.} We generated all such test cases. Running such tests was proposed in previous work by Xie \cite{66} as a method to provide automated crash-testing. The number of obtained test cases varied between target applications. In the case of the FreeMind versions, the number of event-interaction test cases varied between 825 and 9,175. For jEdit, the number of length-2 test cases varied between 3,453 and 9,601.
\item \emph{Randomly generated length 3,4 and 5 test cases.} Generating all length-n, $n > 2$ test cases is not feasible for complex applications. For example, the number of length-3 test cases varied between 12,555 for the simplest version of FreeMind and 323,211 for the most complex jEdit version. Longer sequences increase exponentially in count: for jEdit version 4.3.0final over 300 million length-5 test cases can be generated. Our strategy was therefore to generate 10,000 random test cases for each test case length using GUITAR's \emph{RandomSequenceLengthCoverage} test generator plugin. This allows keeping the number of test cases reasonable while properly sampling the target application's event flow graph.
\end{itemize}

Our strategy resulted in 404,826 FreeMind and 564,869 jEdit test cases that we believe properly sample the target applications' test case space. The next step was to classify generated test cases in one of four categories for all subsequent versions of the target application. The categories were considered to be representative both for state of the art testing tools such as GUITAR and for more advanced implementations that are able to repair or update test cases, such as ones proposed by Memon \cite{97}, Huang \cite{98}, or our own proposed approach \cite{177} based on McMaster and Memon's preliminaries \cite{113}. 

In the following we detail the four categories:

\begin{enumerate}
\item \emph{Replayable using widget Id}. This situation simulates how test cases can be replayed using tools that identify widgets using assigned Id's, an example of which is the GUITAR framework itself. GUITAR components use  widget properties to calculate the \emph{Id}s that are reused when replaying test cases. This approach is generally more accurate than the first wave of capture-replay tools, many of which use positional information to find GUI widgets.
\item \emph{Replayable after repair}. This category is comprised of test cases that can be repaired using previously described approaches \cite{67,68,177} to be replayable on the newer version of the application. This amounts to all GUI elements actioned in the test case, including reaching steps required for enabling further GUI actions to have equivalents on the newer version and their sequence to remain valid according to the newer version's event flow graph. Of course, all test cases that are replayable using widget \emph{Id}s are also replayable using a hypothetical identity repair that does not perform any changes. Test cases in these first categories are exactly replayable on the newer application version provided that highly accurate processes are implemented.
\item \emph{Repairable}. Compared to the category described above, we relax the imposed event flow graph condition and we only require the existence of equivalent widgets on the new application version. Test cases in this category will not have the same sequence of events as the original ones, but they can be repaired using approaches detailed in \cite{97} or \cite{98}.
\item \emph{Unrepairable}. This last category comprises test cases that cannot be repaired. This is due to at least one of the test case widgets missing from the newer application's GUI. To the best of our knowledge, the only way of salvaging these test cases is adding or removing test steps, as detailed by Memon \cite{97}.
\end{enumerate}

Our goal is to evaluate the replayability of GUI test cases on newer versions of the target applications. For this, we considered the generated test cases for all examined versions and categorized them for all subsequent application versions. For example, the 30,824 test cases generated for FreeMind 0.2.0\footnote{The first FreeMind version examined} were categorized for all 12 subsequent examined versions of the application,  the last of which is a September 2007 CVS snapshot of FreeMind.

The present section contains three subsections. The first one details our chosen target applications. The following two subsections present our cross-sectional and longitudinal evaluations. We performed the cross-sectional study to assess the replayability of test suites on consecutive application versions, while the longitudinal approach overviews the results obtained over all the studied versions. In order to limit the effect of randomness in test case generation the procedure was repeated three times with average values obtained reported. 

\subsection{Target applications}

The chosen applications for our case study are two complex, open source GUI-driven applications available under non-restrictive GPL-licences: the FreeMind mind mapper and the jEdit text editor. We chose 30 distinct versions of these applications that range between November 2000 and September 2007 for FreeMind\footnote{13 intermediate versions between 0.2.0 and 0.8.0} and January 2000 to May 2010 for jEdit\footnote{17 versions between 2.3pre2 and 4.3.2final}. Both applications are available on SourceForge\footnote{http://sourceforge.net} where they rank among the most popular applications: FreeMind recorded over 14.3 million downloads over its lifetime while jEdit was downloaded over 6.7 million times since the project was started. Both applications received the \emph{"Project of the Month"} SourceForge award over their lifetimes. With regards to complexity, FreeMind's GUI consists of one main window having between 101 and 280 GUI elements, while jEdit contains between 12 to 16 windows that contain between 482 and 992 GUI elements. Due to limitations of the GUIRipper tool, the \emph{"Options"} window of both applications was not recorded and is disregarded in the present study. A detailed overview of the studied application versions is available in \cite{181}.

\subsection{Cross-Sectional Approach}

The first part of our investigation was undertaken to obtain a detailed picture of GUI test case replayability over targeted application versions. For this, we categorized test cases generated for each version according to their replayability on the application version immediately following it. This approach can isolate application versions for which test cases cannot be repaired and allows corroborating known changes in the GUI with their effect on test case replayability.

\begin{figure}[!t]
\centering
\includegraphics[width=3.5in]{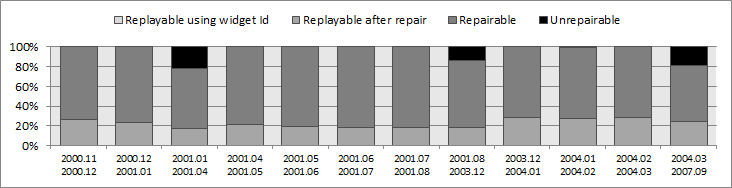}
\caption{Cross-sectional FreeMind test case replayability}
\label{fig:FreeMindCS}
\end{figure}

Figure \ref{fig:FreeMindCS} shows our obtained results for FreeMind. As the application source code was obtained directly from CVS we use timestamps to identify application versions. The first immediate observation is that the number of directly replayable test cases is of no significance. However, we also observe that most test cases can be repaired to run on the newer version of the application. Most unrepairable test cases are found in April 2001, December 2003 and September 2007 versions, which according to our data \cite{181} correspond with major GUI changes.

\begin{figure}[!t]
\centering
\includegraphics[width=3.5in]{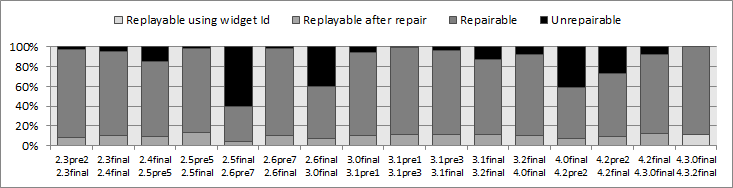}
\caption{Cross-sectional jEdit test case replayability}
\label{fig:jEditCS}
\end{figure}

The corresponding results for jEdit are shown in Figure \ref{fig:jEditCS}. Again we witness most generated test cases being at least repairable across version pairs. However, we must note that several version pairs exhibit large numbers of unrepairable tests. The best example is between versions 2.5final and 2.6pre7, where 60\% of tests become unrepairable. This can be explained using our jEdit GUI model that consists of 16 windows for the former version and 12 for the latter. Tests that target eliminated windows naturally become unrepairable, as the target section of the AUT  no longer exists. 

Taking into account the long time span of the examined versions (7 years for FreeMind and 10 for jEdit), together with jEdit's complex GUI \cite{181}, we conclude that the major factor affecting test case repairability is represented by functional changes in the application GUI. We believe these initial results confirm Memon's previous findings \cite{97} and hint toward the usefulness of efficient approaches for repairing GUI test cases.

\subsection{Longitudinal Approach}

In the second part of our investigation we examine long-term replayability of GUI test cases. For this, we categorize test cases generated for each version on all the subsequent versions of the same application. This provides insight into the long term effects GUI changes have on existing test cases.

\begin{figure*}[!t]
\centering
\includegraphics[width=1.0\textwidth]{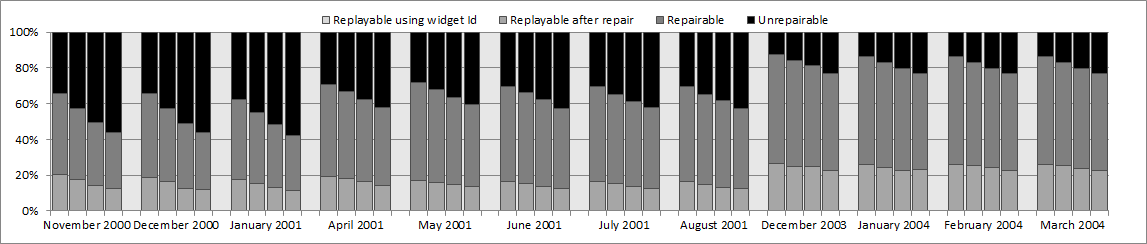}
\caption{Longitudinal FreeMind test case replayability}
\label{fig:FreeMindLong}
\end{figure*}

Figure \ref{fig:FreeMindLong} shows the results of our longitudinal evaluation on the FreeMind application. Each column group shows replayability in the last studied version of FreeMind for test cases generated in the given version. For example, as the last studied version of FreeMind has a timestamp of September 2007, the second column shows the replayability of test cases generated for the December 2000 version on the latest snapshot. As our information only includes data on functionally equivalent widgets in consecutive versions, we had to categorize test cases on each of the intermediary versions up to the final one. In our example, we had to categorize the test suite generated for the December 2000 version on the 10 intermediate versions separating it from the final one. The four columns in each group symbolize test cases of lengths 2,3,4 and 5 in left to right order.

The data confirms our educated guess that the "age" of test cases has an important effect on their replayability. However, we observe that roughly 50\% of all test cases remain at least repairable after 7 years of application development. We attribute this partly to FreeMind's simpler user interface that consists of one window\footnote{From version 0.6.7 it also has an \emph{Options} window that could not be correctly ripped using GUIRipper.} that contains all its widgets. Also, similar to Memon \cite{97}, we find that test case length has a considerable effect on replayability, as longer test cases are more prone to becoming unrepairable. If only one of a test case's events lacks a functional equivalent on the newer version it is immediately categorized as unrepairable. This leads to an interesting issue that may appear when given three consecutive application GUI versions, say $G_1$, $G_2$ and $G_3$. Let us assume that event $e \in G_1$, $e \notin G_2$ but $e \in G_3$, so a GUI event that is removed from an intermediary version but reappears at a later date. This might cause test cases containing $e$, which are unrepairable on $G_2$, to become at least repairable for $G_3$. As generally no information is available regarding an application's \emph{future} GUI structure, we consider unrepairable test cases to remain as such. This also holds for the other two categories employed.

\begin{figure*}[!t]
\centering
\includegraphics[width=1.0\textwidth]{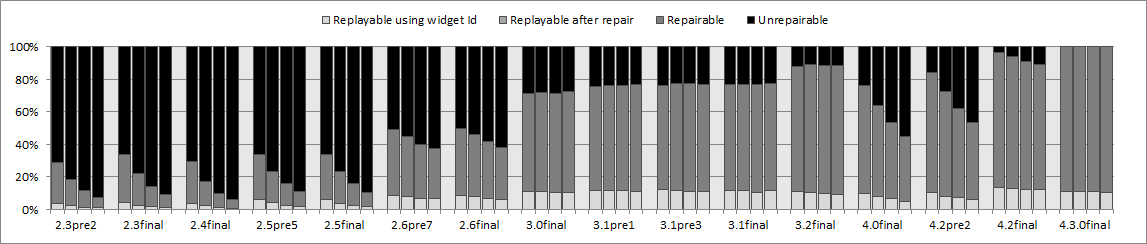}
\caption{Longitudinal jEdit test case replayability}
\label{fig:jEditLong}
\end{figure*}

Figure \ref{fig:jEditLong} provides the longitudinal data for the jEdit application. Our observations for FreeMind generally hold in the case of our second application: older or longer test cases have higher probability of not being reusable. An interesting aspect regards the result of jEdit's GUI model changes between versions 2.5final and 2.6pre7. In the previous section we witness GUI changes between the versions making roughly 60\% of test cases unreplayable on the latter version. In the longitudinal view the specified version pair acts as a choke-point, as we witness most test cases generated prior to version 2.6pre7 being unusable on our latest studied version, 4.3.2final, due to them being broken by changes in version 2.6pre7. This enforces our belief that major GUI changes are a serious hurdle in automating GUI regression testing, and processes based on research such as detailed in \cite{97,98,177} must take steps to alleviate these issues. Also, our evaluation shows that investing effort in processes related to test case maintenance is worthwhile, as a highly accurate automated process will be able to consistently repair old test cases to work on new versions of the AUT, even with long timespans considered.

\section{Threats to validity}

We partition threats to the validity of our empirical evaluation into internal and external. Internal threats are represented by errors in the process employed to obtain our data. In this regard, the main issue regards the fact that our evaluation was performed by \emph{simulating} test case execution using test case data together with GUI and event flow graph models. This enabled us to generate and categorize over 24 million test cases, including ones generated for intermediate versions for the longitudinal study. However, this has the drawback that errors in recorded GUI models or implementation peculiarities of GUITAR components or the applications themselves might cause test cases to belong to other categories than those assigned to in our study.

External threats regard the generalization of obtained information. While FreeMind and jEdit are complex real-life applications having extensive user-bases and participating in previous empirical research \cite{145,101,97,67,66}, they are not representative of all possible GUI implementations. Practitioners looking to capitalize on our results must have a good understanding on the presented limitations and peculiar aspects regarding their targeted applications.

Our best effort to mitigate presented threats is to make all our data available for analysis on our website \cite{105}. This includes the source code that categorizes test cases, the GUI models employed together with test case information.

\section{Conclusions and Future Work}

In this paper we presented an initial study on what can be expected in long term GUI test case replayability in the case of complex open source software. We performed a cross-sectional evaluation where generated test cases were replayed on the immediately following studied version of the target application. We used this information to detail the results of the longitudinal evaluation where we performed a simulation of long term replayability for GUI test cases. We believe our results validate previous work in GUI test case maintenance \cite{97,98,113,177} and we hope to fuel further work in the field.

Due to our promising initial results, our next goal is to conduct a comprehensive follow-up study employing a larger selection of target applications on other platforms such as .NET and SWT. This entails extending our software repository with new application versions and obtaining associated information regarding functionally equivalent GUI widgets. In addition, we aim to switch from \emph{simulating} test case execution to running them by employing GUITAR's test runner component, thus eliminating one of the threats to the validity of the presented research.

A more distant avenue of research regards a comprehensive evaluation targeting event-driven systems beyond the desktop paradigm by including web and mobile applications. Theoretical advances in unified modelling of event driven software \cite{168} together with GUITAR components targeting these platforms enable such a complex undertaking. We believe such an effort can lead to better understanding of GUI-driven software in a platform independent manner and enable the creation of unified testable models for such systems.

\section*{Acknowledgment}

The author was supported by programs co-financed by The Sectoral Operational Programme Human Resources Development, Contract POS DRU 6/1.5/S/3 - ``Doctoral studies: through science towards society''

\bibliographystyle{IEEEtran}
\bibliography{IEEEabrv,biblio}

\end{document}